# Analytical Modelling of Ferroelectricity Instigated Enhanced Electrostatic Control in Short-Channel FinFETs

Jhang-Yan Ciou[1], Sourav De[1, *], Chien-Wei-Wang[1], Wallace Lin[1], Yao-Jen Lee[2], and Darsen Lu[1, *]

[1]Institute of Microelectronics, Department of Electrical Engineering, National Cheng Kung University, Tainan, Taiwan. [2]Taiwan Semiconductor Research Institute, Hsinchu, Taiwan.

e-mail: q18077502@gs.ncku.edu,tw, darsenlu@mail.ncku.edu.tw

*Abstract*— This study simulated negative-capacitance double gate FinFETs with channel lengths ranging from 25nm to 100nm using TCAD. The results show that negative capacitance significantly reduces subthreshold swing as well as drain induced barrier lowering effects. The improvement is found to be significantly more prominent for short channel devices than long ones, which demonstrates the tremendous advantage of negative capacitance gate stack for scaled MOSFETs. A compact analytical formulation is developed to quantify sub-threshold swing improvement for short channel devices.

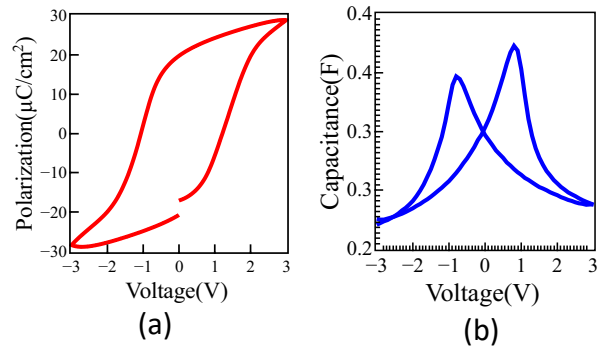

Fig. 1. (a) Measured polarization vs voltage(P-V) curve. The measurement was conducted using Radiant Precision LC meaurement setup. The small opening in P-V curve depicts neglible RC leakage and (b) corresponding capacitance vs voltage(C-V) curve, measured by B1500 semiconductor device analyzer, shows the position of coercive voltage around 1V. The peaks in the C-V curve manifests the presence of ferroelectricity in the fabricated film. The assyemtry in peak heights in C-V curve is the aftermath of assymetric top(TiN) and bottom electrode(TiN-SiO$_2$-Si). However, this offset was balanced during the P-V measurement.

## I. INTRODUCTION

As the field-effect transistor's (FET) physical channel length scales down, both the switching speed and the drive current are improved at lower power consumption. However, at sub-100nm technology nodes, we face tremendous challenges related to the trade-offs between on-state drive current ($I_{on}$) and power consumption due to degraded short channel effects (SCE). In order to simultaneously reduce power consumption and maintain $I_{on}$, the negative capacitance FET (NCFET) was proposed [1, 2]. It replaces conventional gate dielectrics with one that includes a ferroelectric (FE) material layer, which provides negative differential capacitance over a wide voltage range. NCFETs exhibit amplification effects from the gate voltage to the surface potential, thereby increasing the amount of inversion charge in the channel. The theory of operation for NCFET had been researched extensively [1-5]. In this paper, we focus on the NCFET's short channel effects. We simulate double-gate (DG) NC FinFET with channel length ranging from 25nm to 100nm. We show that as the channel length is reduced, the capacitive coupling from the drain assists the matching between normal capacitances and ferroelectric (negative) capacitance, which consequently improves the performance (sub-threshold swing) of DG-NC-FinFET.

## II. FABRICATION AND CHARACTERIZATION OF HZO BASED FERROELECTRIC CAPACITORS

We have fabricated the ferroelectric capacitors with 10nm thick HZO. The experimental data obtained from the fabricated devices are used for conducting TCAD simulation of negative capacitance Fin-FET (NC-FinFET) using Preisach model of hysteresis. Fig. 1(a) shows the measured polarization vs applied voltage curve (P-V) and Fig. 1(b) shows the corresponding capacitance vs voltage (C-V) curve. The extracted values of coercive voltage, remnant polarization and saturation polarization were used for conducting transient TCAD simulation to unravel the capacitance response of HZO with changing electric field.

## III. DEVICE STRUCTURE AND SIMULATION METHOD

Figs. 2(a) and (b) show the simulated structure for DG-NC FinFET along with gate-stack and geometry parameters. The structure and device designs are adopted from [4,5]. The thickness of the interfacial oxide and FE layer are 1nm and 3nm, respectively. The FE layer's intrinsic characteristics are given in the form of coercive field ($E_c = 1.2 \times 10^6$ V/cm) and remnant polarization ($P_r = 1.5 \times 10^{-5}$ C/cm$^2$.) [6], which can be extracted from the *P-E* curve in Fig. 1. TCAD [8] was used to simulate device characteristics for a given device structure using the aforementioned parameters. With TCAD, the FE polarization mechanism can be switched on or off. The FE layer exhibits NC characteristics only when the ferroelectric polarization mechanism is turned on. Otherwise, the FE layer exhibits normal dielectric characteristics. In order to make a fair comparison, our experimental group (NC-FinFET) and control group (conventional FinFET) have the same device structure, gate stack, simulation mesh, and material parameters, etc., except that ferroelectric polarization mechanism is turned off for the control group.

The simulation process is performed as follows. The FE capacitance ($C_{FE}$) is verified to be negative first. This is followed by modulation of the interfacial oxide layer thickness at a fixed FE thickness. Finally, $C_{FE}$ is computed with extrapolation.

To account for hysteresis effects in double gate NC-FinFETs, we perform transient TCAD simulation with channel length varying from 25nm to 100nm. Device characteristics of subthreshold swing, on current, and off current are then extracted subsequently.

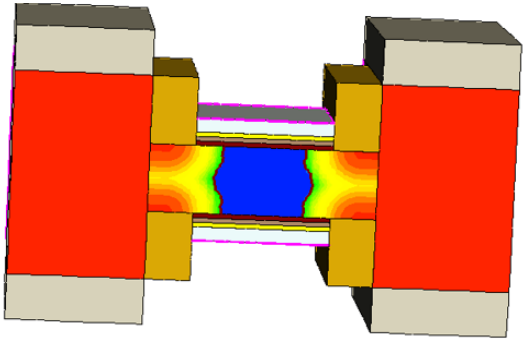

(a)

**Device Parameters**

| Gate Length ($L_c$) | 25nm–100nm |
|---|---|
| Fin Width ($T_{fin}$) | 10nm |
| Fin Height ($H_{fin}$) | 30nm |
| IL Oxide Layer Thickness ($T_{ox}$) | 1nm |
| FE Layer Thickness ($T_{FE}$) | 3nm |

(b)

Fig. 2. (a) Three-dimensional NC FinFET structure studied in simulation. (b) List of nominal device parameters used in TCAD simulation.

## IV. RESULT AND DISCUSSION

Fig. 3(a) illustrates the equivalent circuit of the ferroelectric gate-stack structure in a double-gate FinFET with a very thin body. Since the thin body region is typically fully depleted, $C_{dep}$ was assumed to be zero. Thus, the total gate capacitance ($C_{total}$) is modeled with the following expression:

$$\frac{1}{C_{total}} = \frac{1}{C_{FE}} + \frac{1}{C_{MOS}} \quad (1)$$

In addition to verifying that $C_{FE}$ is negative, we would also like to obtain its exact value. Therefore, a simulation experiment specifically for extracting $C_{FE}$ was designed according to (1). $C_{FE}$ was obtained by extrapolation of multiple data points of $C_{total}$ and $C_{MOS}$. C-V simulation was performed under the condition that the DG NC FinFETs have relatively long channel length (100nm). We fix the FE layer thickness while varying the interfacial oxide layer's thickness ($T_{ox}$) to be 1nm, 3nm and 4nm respectively. Fig. 3(b) shows the resulting C-V characteristics. $C_{total}$ is larger than $C_{ox}$ in all three cases. This is due to the fact that $C_{ox}$ series combination with the negative $C_{FE}$ actually increases $C_{total}$. $C_{FE}$ is then extracted based on Eq. (1), as shown in Fig. 2(c). The extracted value for $C_{FE}$ is $-6.2554 \times 10^{-17}$ F at 1V. Many researchers have pointed out that negative-capacitance gate stacks may help conventional FET devices improve subthreshold swing (SS), on-current, and reduce off-current and threshold voltage ($V_{th}$) roll-off [1-5]. At present, fabrication of uniform ferroelectric films down to a very short channel length is difficult due to grain size and other limitations. As a result, TCAD simulation is explored to predict ultimate scaling behavior of negative capacitance DG FinFET. Note that the extracted SS in Fig. 4 is an average value over a voltage range from 0.16 to 0.2V. The long-channel SS value is also extracted at channel length 100nm showing SS around 58.8 mV/dec. Fig. 4 illustrates the trend that the SS improvement becomes more significant as the channel length become shorter. For instance, The SS improvement at channel length of 25 nm is as much as 54.67%. Such observation is in line with that reported in [9].

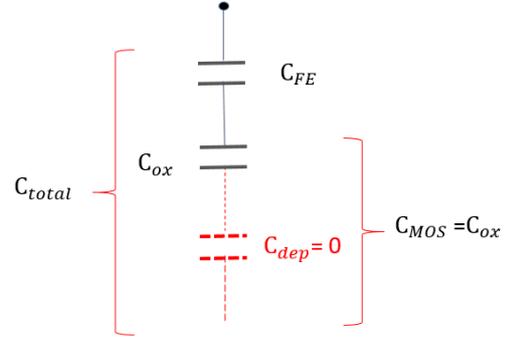

(a)

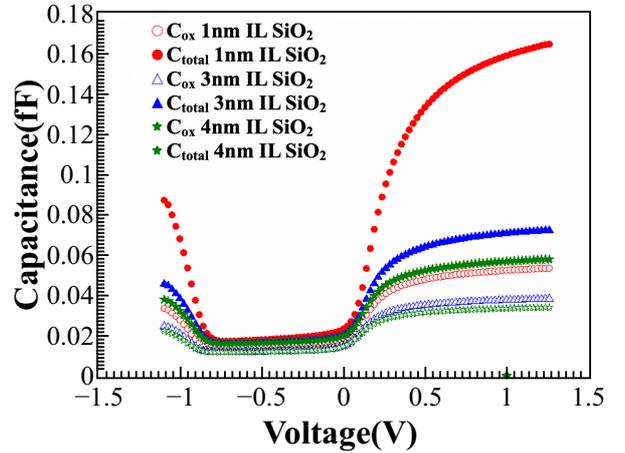

(b)

Fig. 3. Equivalent circuit of the gate stack where $C_{total}$ represents the total gate capacitance including $C_{FE}$. (b) C-V simulation result of gate stacks varying with three different interfacial oxide thicknesses. The solid symbols are C-V with negative capacitance effect; the hollow symbols are C-V of the control group where ferroelectric polarization is intentionally switched off in TCAD simulation.

We would like to further develop an analytical model to quantitatively explain the above phenomenon. The NC gate-stack amplify the gate voltage to improve device performance. This amplification ratio is given by [10]:

$$A_V = \frac{\partial V_{MOS}}{\partial V_G} = \frac{|C_{FE}|}{|C_{FE}| - C_{MOS}(C_d)} \quad (2)$$

The value of $C_{MOS}$ is influenced by the drain induced barrier lowering (DIBL) effect, which is translated to SS improvement. The focus in [10] is on drain biasing effect, whereas we focus on channel length dependence in this study.

The effect of drain-side electric field coupling can be modelled with a DIBL capacitance, $C_d$, which degrades conventional FET device performance. On the other hand, $C_d$ may improve NC gate-stack capacitance matching (Fig. 5). In mathematically terms, the DIBL capacitance is expressed as follows, based on a simplified version of BSIM-CMG sub-threshold swing degradation model [11]:

$$C_d = \frac{0.5}{\cosh\left(DVT\frac{L_c}{\lambda}\right)-1}(C_{DSC}) \quad (3)$$

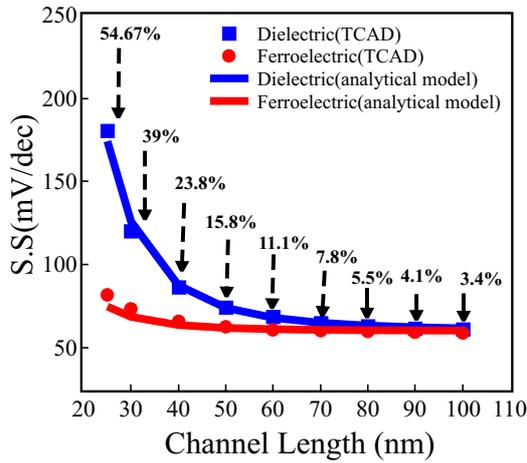

Fig. 4. Sub-threshold swing versus channel length, with and without turning on ferroelectric polarization effects in TCAD simulation. The percentage decrease of subthreshold swing is depicted by orange dashed arrows. Calibration result for SS as a function of channel length. Solid lines are model prediction.

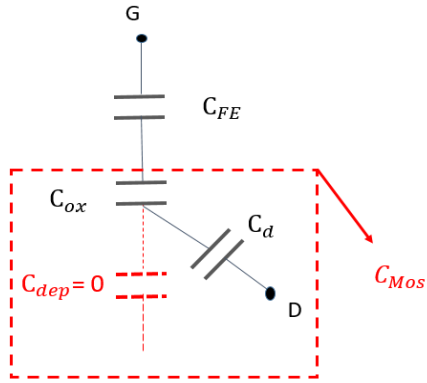

Fig. 5. Simple equivalent capacitance network model to take into account DIBL effect.

As $L_c$ scales down, according to Eq. (3), DIBL effect becomes more severe since $C_d$ increases. This leads to a larger value of $C_{Mos}$. Consequently, the $C_{Mos}$ value would approach $|C_{FE}|$. As a result, the denominator in Eq. (2) becomes smaller, and $A_V$ increases. The larger $A_V$ also translates to better SS, as can be seen from the following expressions for the relationship between SS and $A_V$ [10]:

$$SS_{FinFET} = 60 mv/dec \left(1 + \frac{C_d}{C_{ox}}\right) \quad (4)$$

$$SS_{NC-FinFET} = SS_{FinFET} \times \frac{\partial V_G}{\partial V_{MOS}} = SS_{FinFET} \times A_V^{-1} \quad (5)$$

According to Eqs. (4) and (5), the SS improvement for NC-FinFETs over regular FinFETs is determined by $A_V$. This explains the more significant SS improvement at short $L_c$ where $A_v$ becomes larger due to larger drain coupling capacitance, $C_d$. The parameters DVT and CDSC in Eq. (3) are calibrated through fitting to TCAD-simulated SS v.s. $L_c$ characteristics, as shown in Fig. 4. The value of DVT is found to be $6.9999 \times 10^{-2}$, whereas CDSC is $1.292 \times 10^{-18}$ F. Eq. (5) is able to accurately predict SS degradation with and without FE effects. This validates the correctness of our analytical model and sub-circuit concept. It is important to note that according to Eq. (2), the improvement of SS is based on the assumption that $C_{MOS}$ does not exceed $|C_{FE}|$, which is true throughout this study.

CONCLUSION

This study has successfully demonstrated using TCAD simulation and analytical modeling to achieve the transistor electrostatics improvement via gate voltage amplification from ferroelectric gate stack. Additionally, the sub-threshold swing improvement for short channel FinFETs is found to be more significant than long channel ones, due to improved capacitance matching by the larger drain-side coupling capacitor. Therefore, drain-side coupling capacitance and its role in matching with $C_{FE}$ must be taken into account during NC-FinFET device design.

ACKNOWLEDGEMENT

This research work was supported by Ministry of Science and Technology (Taiwan) grant MOST-107-2634-F-006-008. The authors would like to acknowledge National Center for High Performance Computing for providing TCAD floating license.